\documentclass{pasa}%

\usepackage{graphicx}
\usepackage{physics}
\usepackage[separate-uncertainty=true]{siunitx}
\usepackage{booktabs}
\usepackage{nicefrac}
\usepackage{subcaption}
\usepackage[normalem]{ulem}

\DeclareSIUnit\beam{beam}
\DeclareSIUnit\jansky{Jy}
\DeclareSIUnit\parsec{pc}
\DeclareSIUnit\solarmass{M_\odot}
\DeclareSIUnit\deg{deg}
\DeclareSIUnit\pixel{pixel}
\DeclareSIUnit\gauss{G}

\title[Stacking with FIGARO]{Stacking the Synchrotron Cosmic Web with FIGARO}

\author[Hodgson et al.]{Torrance Hodgson$^{1,2}$\thanks{\href{mailto:torrance@pravic.xyz}{torrance@pravic.xyz}}, Franco Vazza$^{3,4,5}$, Melanie Johnston-Hollitt$^{1,2}$ and Benjamin McKinley$^{1,6}$
\affil{$^1$International Centre for Radio Astronomy Research (ICRAR), Curtin University, 1 Turner Ave, Bentley, 6102, WA, Australia}
\affil{$^2$Curtin Institute for Computation, Curtin University, GPO Box U1987, Perth, 6845, WA, Australia}
\affil{$^3$Dipartimento di Fisica e Astronomia, Universit\a'a di Bologna, Via Gobetti 92/3, 40121, Bologna, Italy}
\affil{$^4$Hamburger Sternwarte, Gojenbergsweg 112, 21029 Hamburg, Germany}
\affil{$^5$INAF, Istituto di Radio Astronomia di Bologna, Via Gobetti 101, 40129 Bologna, Italy}
\affil{$^6$ARC Centre of Excellence for All Sky Astrophysics in 3 Dimensions (ASTRO3D), Bentley, Australia}
}%

\jid{PASA}
\doi{10.1017/pas.\the\year.xxx}
\jyear{\the\year}

\usepackage{aas_macros}
\usepackage{hyperref} 
\hypersetup{colorlinks,citecolor=blue,linkcolor=blue,urlcolor=blue}

\begin{document}

\begin{frontmatter}
\maketitle

\begin{abstract}
Recently \citet{Vernstrom2021} claimed the first definitive detection of the synchrotron cosmic web, obtained by `stacking' hundreds of thousands of pairs of close-proximity clusters in low-frequency radio observations and looking for a residual excess signal spanning the intracluster bridge. A reproduction study by \citet{Hodgson2022},  using both the original radio data as well as new observations with the Murchison Widefield Array, failed to confirm these findings. Whilst the detection remains unsure, we here turn to stacking a \textit{simulated} radio sky to understand what kind of excess radio signal is predicted by our current best cosmological models of the synchrotron cosmic web. We use the FIlaments \& GAlactic RadiO \citep[FIGARO;][]{Hodgson2021} simulation, which models both the synchrotron cosmic web as well as various subtypes of active galactic nucleii and star forming galaxies. Being a simulation, we have perfect knowledge of the location of clusters and galaxy groups which we use in our own stacking experiment. Whilst we do find an excess radio signature in our stacks that is attributable to the synchrotron cosmic web, its distribution is very different to that found by \citet{Vernstrom2021}. Instead, we observe the appearance of excess emission on the immediate interiors of cluster pairs as a result of asymmetric, `radio relic'-like shocks surrounding cluster cores, whilst the excess emission spanning the intracluster region---attributable to filaments proper---is two orders of magnitude lower and undetectable in our experiment even under ideal conditions. 

\end{abstract}

\begin{keywords}
Cosmic web (330) -- Warm-hot intergalactic medium (1786) -- Radio astronomy (1338)
\end{keywords}
\end{frontmatter}

\section{Introduction}

The imagery of the `cosmic web' has been used to describe the Universe on the very largest of scales. It describes the ongoing process of structure formation, starting with primordial perturbations in the mass distribution of the early Universe, and since growing into an ontology of structures: large-scale voids, emptying onto surrounding sheets, collapsing down into filaments, and feeding into galaxy groups and clusters. Empirical evidence of this structure has come from galaxy surveys \citep[e.g.][]{Baugh2004}, but cosmological simulations also suggest that these structures are much more massive than the galaxies that trace them out \citep[e.g.][]{Cen1999, Dave2001}, with up to 40\% of the baryon population of the Universe---the so-called `missing baryon' population \citep[e.g.][]{Nicastro2017}---located along the filaments and around the periphery of clusters. Empirical corroboration has proved difficult since the majority of the matter that traces this structure is predicted to exist in a low density, warm-hot (\SIrange[range-phrase=--,range-units=single]{E5}{E7}{\kelvin}) and highly ionised state, rendering it extremely difficult to detect in practice.

Nonetheless, numerous reports have claimed tentative detection of this `warm-hot intergalactic medium' (WHIM), backing up the simulation predictions. \citet{Eckert2015}, \citet{Nicastro2018}, \citet{deGraaff2019}, \citet{Tanimura2019} and \cite{Tanimura2020}, for example, used a range of techniques such as molecular absorption lines and statistical (stacking) detections of the Sunyaev-Zeldovich effect to claim detections of this large scale structure. Perhaps the most definitive accounting for the missing baryon population was recently made by \citet{Macquart2020} using fast radio bursts to trace the intervening density of the Universe, and finding it to be overdense consistent with the missing baryons residing, hidden, along the line of sight.

Simulations have also pointed to the existence of a radio synchrotron component that traces out the cosmic web \citep[e.g.][]{Brown2011,ArayMelo2012,Vazza2015,Vazza2019}. This hypothesised emission is driven by accretion processes occurring within the WHIM---as part of the ongoing large-scale structure formation of the Universe---that are expected to produce strong shocks, with Mach numbers in the range $\mathcal{M}\sim$~\SIrange[range-phrase=--,range-units=single]{10}{100}{}. These shocks occur in the low density peripheries of clusters and around filaments, and they are capable of accelerating a small fraction of the ambient electron population to relativistic energies by way of diffusive shock acceleration processes \citep[e.g.][]{Keshet2009}. In the presence of intracluster magnetic fields, which we expect to be on the order of a few nG \citep[e.g.][]{Pshirkov2016,OSullivan2019,Vernstrom2019,Carretii2022}, these energetic electron populations should radiate their energy as synchrotron emission. But their emission is extremely faint. Simulations by \citet{Vazza2015}, for example, predict only the very brightest peaks of emission from the synchrotron cosmic web to be detectable with the current generation of radio telescopes, and detection is made more difficult still due to much more luminous radio populations such as active galactic nuclei (AGNs) or star forming galaxies (SFGs).

The difficulty of direct detection has driven statistical detection techniques, and foremost among these has been the cross-correlation method. This method seeks to overcome both the problem of the faint signal of the synchrotron cosmic web and the obscuring effect of the more luminous radio source populations by essentially integrating across a large area of sky. It does this by identifying a `best guess' distribution of the synchrotron cosmic web in a region of the sky, and performing a radial cross-correlation of this kernel with the observed radio emission. In theory, a peak at or near \SI{0}{\degree} offset would indicate a detection of the cosmic web. In practice, however, lots of other emission sources cluster similarly and pollute the signal \citep{Hodgson2021}. And so while the cross-correlation analysis of \citet{Vernstrom2017} did indeed detect a peak at \SI{0}{\degree}, they were unable for this reason to make any kind of definitive claim of detection. Meanwhile, a similar cross-correlation analysis by \citet{Brown2017} found no detection at all.

Recently, however, there was a notable development in this field with the announcement of the radio detection of filaments by \citet{Vernstrom2021}, herein V2021. This was also a statistical detection, but instead the authors employed a stacking technique. Their method used luminous red galaxies (LRGs) derived from the Sloan Digital Sky Survey Data Release 5 \citep{Lopes2007} as tracers for overdense regions such as clusters and galaxy groups. These LRGs had photometric redshift data, and so it was possible to create a catalogue of LRG pairs separated by no more than \SI{15}{\mega \parsec}, with the assumption that close proximity clusters will, on average, be connected by filaments. V2021 proceeded by stacking continuum images from two low frequency radio surveys, the GaLactic and Extragalactic All-sky MWA\footnote{Murchison Widefield Array \citep{Tingay2013}.} survey \citep[GLEAM;][]{Wayth2015,HurleyWalker2017} and the Owens Valley Radio Observatory Long Wavelength Array \citep[OVRO-LWA;][]{Eastwood2018}. Pairs of LRGs were extracted from these sky surveys, rotated, rescaled, and stacked along a normalised coordinate frame such that each pair was positioned at $x = -1$ and 1, respectively. After many such pairs were stacked, the expectation was that the image noise would be sufficiently reduced so that excess filamentary emission would become detectable along the bridge between -1 and 1. Indeed, V2021 reported excess filamentary emission along this bridge at \SI{118}{\mega \hertz} with a temperature of \SI{0.22}{\kelvin} and having a spectral index of $\alpha = -1.0$. Moreover, a null test with spatially distant pairs of LRGs produced no excess emission. The result of V2021 is surprising for a number of reasons, not least in that it implies intracluster magnetic field strengths that are stronger than previous upper limits, in the range of \SIrange[range-phrase=--,range-units=single]{30}{60}{\nano \gauss} for a significant fraction of filaments. In fact, we should note this is strictly a lower limit, as not all LRG pairs are centrally aligned with a host cluster or galaxy group, and further only a fraction of these will in fact be connected by filament.

Given this surprising result, \citet{Hodgson2022}, herein H2022, attempted to reproduce this result by stacking both the original GLEAM survey data as well as stacking new observations made with Phase II of the MWA \citep{Wayth2018}. Their stacking methodology closely followed that described in V2021, and stacked the same catalogue of LRG pairs as well as additional, differently parameterised catalogues of LRG pairs. In each case, no excess emission was detected that resembled the broad excess intracluster emission as detected in V2021.

Whilst it remains unclear the cause for this discrepancy in results, we can meanwhile turn to cosmological simulations to ask what kind stacking profile we should expect. In this paper, we make use of the FIlaments and GAlactic RadiO simulation \citep[FIGARO;][]{Hodgson2021} to perform our own simulated stacking experiment. FIGARO simulates the low to mid frequency radio sky including AGN and SFG populations, against a backdrop of large-scale diffuse cosmic web synchrotron emission produced by \citet{Vazza2019}, the largest magneto-hydrodynamic (MHD) cosmological simulation to date. Crucially, these different radio populations coherently cluster together with respect to an underlying mass distribution drawn from the same cosmological volume. FIGARO is able to produce light cones of each of these populations for both configurable fields of view and variable redshift depths. 

We use FIGARO to create ten, \SI{15 x 15}{\degree} fields at \SI{150}{\mega \hertz} using an observing configuration designed to match the MWA Phase I instrument \citep{Tingay2013}. We aim to reproduce the stacking methodology of V2021 and H2022 applied to these simulated radio fields and thus construct simulated stacked profiles of both the synchrotron cosmic web as well as the much brighter embedded radio populations of AGN and SFG. The primary deviation from their methodology, however, is that we do not need to use LRGs as a proxy for cluster locations, since we have perfect knowledge of the location of dark matter (DM) halos throughout our fields, and we can thus stack clusters and galaxy groups directly. We hope by doing this exercise to give a sense of the expected magnitude and distribution of excess emission signature expected from any future stacking experiments.

This paper begins by outlining the construction of the \SI{15 x 15}{\degree} simulated fields in Section 2. We follow in Section 3 by outlining the stacking process, which includes our choice of DM halo pairs. Finally, in Section 4 we discuss the results of our stacks, our predictions for future, deeper stacking experiments as well as the challenges of the technique.

Throughout this paper we assume a $\Lambda$CDM cosmological model, with density parameters $\Omega_{\text{BM}} = 0.0478$ (baryonic matter), $\Omega_{\text{DM}} = 0.2602$ (dark matter), and $\Omega_{\Lambda} = 0.692$, and the Hubble constant $H_0 =$ \SI{67.8}{\kilo \metre \per \second \per \mega \parsec}. These values are consistent with those used in both FIGARO and, ultimately, the original simulation by \citet{Vazza2019}.

\section{Field construction}

The first step in this exercise is to produce the simulated fields that we will ultimately use in our stacking experiment. To produce these fields, we have made use of the FIGARO simulation, which allowed us to construct realistic maps of the radio sky incorporating not just the synchrotron cosmic web, but also AGN and SFG populations. FIGARO is built off an underlying $100^3$~\SI{}{\mega \parsec \cubed} MHD cosmological simulation from \citet{Vazza2019} which, for our purposes, allows us to track the evolution of mass density, accretion shocks, and the resulting radio synchrotron cosmic web emission. To this, AGN and SFG populations were added in accordance to the underlying mass distribution using T-RECS simulation software \citep{Bonaldi2019}, ultimately allowing us to create a realistic radio sky where mass, extragalactic radio sources, and the synchrotron cosmic web are distributed and cluster accurately with regards to one another. The full details of FIGARO, including its calibration of cosmic web emission to the observed radio relic population, are provided in \citet{Hodgson2021}.


We have used FIGARO to construct ten \SI{15x15}{\degree} fields---or `realisations'---out to a redshift depth of $z = 0.6$ and at an observing frequency of \SI{150}{\mega \hertz}. As with the original FIGARO, each realisation differs by laterally offsetting and rotating the simulation volume by some random factor each time it is appended in redshift space. For low redshifts, each realization can appear quite drastically different depending on whether, by chance, the lateral offset places a massive cluster in the near foreground, or perhaps instead a massive void. By using ten such realizations, we hope to give a good sense of near-field cosmic variance. With a field this large, however, it becomes necessary to laterally replicate the underlying cosmological volume for redshifts $z > 0.09$, since the field of view spans greater than the \SI{100}{\mega \parsec} width of the underlying cosmological volume. The result of this is that there will be increasing redundancy of any stacking procedure as we go deeper in redshift space, since the same cosmic web features will be repeated more than once across the field of view. In practice, however, this increasing redundancy at high redshifts is mitigated by our selection criteria for halo pairs (see below) as we make use of a much more local region of redshift space, with more than half of the DM halo pairs located within $z < 0.2$. Moreover, whilst a cosmic web feature may appear more than once across a redshift slice, each will be obscured by a unique foreground of AGN, SFG and unrelated cosmic web emission. The higher redshift cosmic web emission, meanwhile, serves primarily as a potentially obfuscating background to this foreground emission, just as we'd expect in practical observations of the radio sky.

For each of these realizations, FIGARO produces a catalogue of AGNs, SFGs, and DM halos, as well as a map incorporating the flux sum of cosmic web emission along the line of sight (in units \SI{}{\jansky \per \pixel}).

\begin{figure}
    \centering
    \includegraphics[width=\linewidth,clip,trim={0 0 1.2cm 0}]{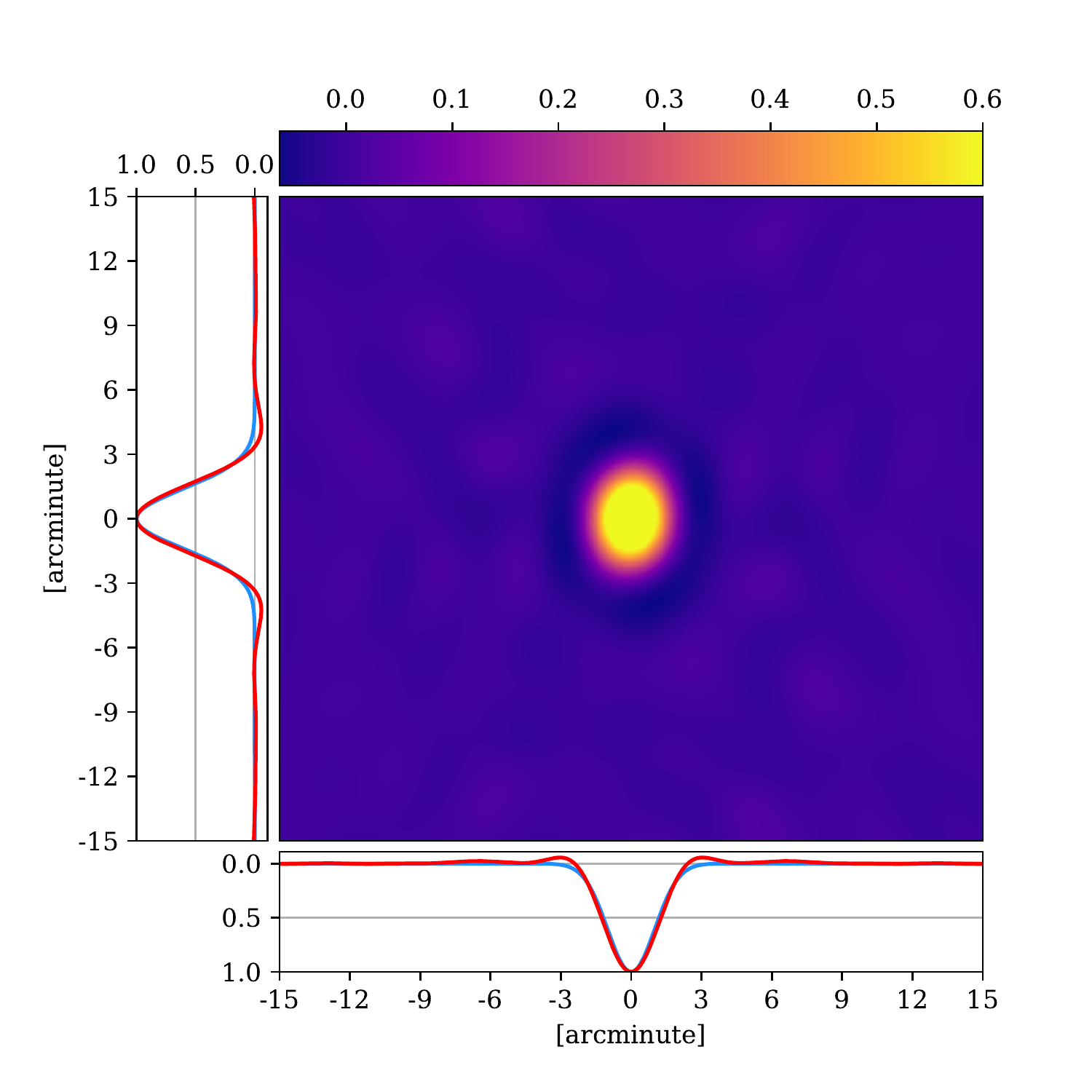}
    \caption{The dirty beam used in this modelling, which closely approximates the Phase I \SI{154}{\mega \hertz} MWA dirty beam at $\sim$\SI{19}{\degree} declination and 30 minutes integration time. The red lines indicate the 2D profile of the dirty beam and the blue lines indicate the elliptical Gaussian fit with parameters \SI{196.4 x 141.6}{\arcsecond} and position angle \SI{-14.5}{\degree}.}
    \label{fig:psf}
\end{figure}

The next step is to transform the FIGARO catalogue into a map of the sky. In an effort to approximate the results of V2021 and H2022, we produce this map at the resolution of the MWA Phase I instrument. We do this by first constructing a map of the sky, in units \SI{}{\jansky \per \pixel}, where the pixel scale is sufficiently smaller than the MWA Phase I point spread function (PSF). FIGARO already provides us with such a map for the cosmic web, which we have set to a pixel scale of \SI{4 x 4}{\arcsecond}. And in the case of the AGN and SFG populations, we model these as simple point sources---that is, single pixel values---which is a good approximation given the low resolution of the MWA Phase I instrument. The final step is to convolve this map with the MWA Phase I PSF, transforming the units from \SI{}{\jansky \per \pixel} to \SI{}{\jansky \per \beam}. At no point do we add simulated thermal or sky noise to our images.

To do this final step, we need to model the MWA Phase I PSF. We have used a sample of archival Phase I observations from the original GLEAM survey to accurately reconstruct the shape of the PSF based on the geometry of the array, elevation of the pointings, and weighting of the baselines. Note that these archival observations are centred with a declination of $\delta =$ \SI{18.6}{\degree}, corresponding to the necessary low elevation pointings used in both V2021 and H2022 to observe the LRG population. These low elevation pointings result in a fairly significant lengthening of the PSF, and it is important in this simulated stacking that we replicate this elongated elliptical beam. Using the same process described in H2022, we find a PSF as shown in shown in \autoref{fig:psf}. When fitted with an elliptical Gaussian, this PSF has dimensions \SI{196.4 x 141.6}{\arcsecond}, with position angle \SI{-14.5}{\degree}. For simplicity, we use this this fitted Gaussian beam when constructing our fields.

\begin{figure*}
    \centering
    \includegraphics[width=\linewidth,clip,trim={1cm 0 0.5cm 0}]{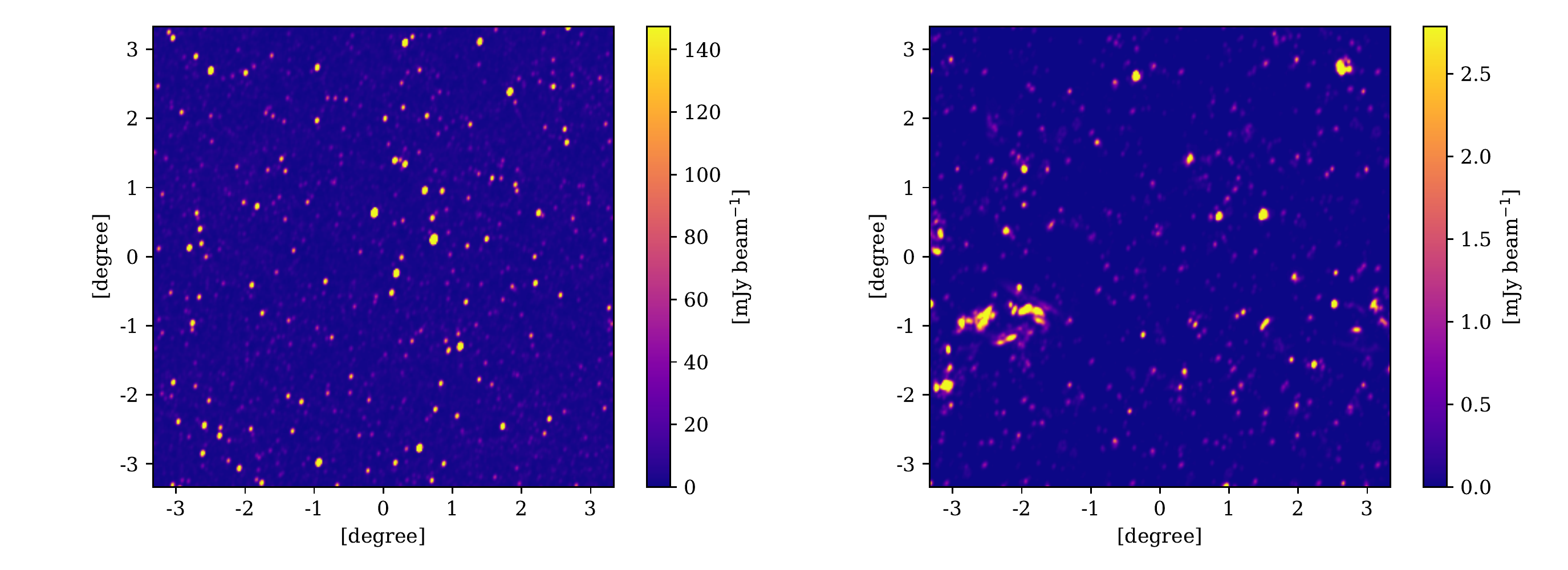}
    \caption{A $\sim$\SI{6 x 6}{\degree} subregion of the field 1, having been convolved with a Gaussian beam approximating the MWA Phase I beam. The colour scale for each map has been capped at the 99.5th percentile pixel value. \textit{Left:} The AGN and SFG map, with a bright \SI{30}{\jansky} source located near the centre. Prior to stacking, this is cleaned down to \SI{10}{\milli \jansky \per \beam}. \textit{Right:} The cosmic web map, showing some nearby, extended emission structures in the bottom left.}
    \label{fig:allfields}
\end{figure*}

With these steps complete, we produce sky maps for each realization. As an example, in \autoref{fig:allfields} we show the inner region of realization 1. On the left of the figure, we show the AGN and SFG population after having been convolved with the fitted Gaussian beam, and on the right the synchrotron cosmic web having been convolved to the resolution of the same beam. For each realization we produce maps of: the AGN and SFG populations; the cosmic web emission; and a third combined map.

\subsection{Point source subtraction}

Both V2021 and H2022 perform a point source subtraction step, with the aim to avoid contamination from comparatively bright AGN and SFG sources during stacking. In V2021, wavelet-based point source subtraction was performed upon GLEAM images, down to a threshold of 5 times the image noise, corresponding to about \SI{120}{\milli \jansky \per \beam}. H2022 instead used the residuals after cleaning, removing all emission brighter than approximately \SI{60}{\milli \jansky \per \beam}, with the assumption that bright emission---point like or not---was unlikely to be attributable to filamentary emission.

We employ this latter cleaning technique to simply and effectively remove bright AGN and SFG emission. The cleaning of our dirty images was performed using a simple image-based algorithm, the equivalent of the Hogbom clean algorithm \citep{Hogbom1974} or the purely `minor' cycles of the Clark algorithm \citep{Clark1980}. The gain parameter was set at 0.1 and the process was continued for each image until no peaks remained above a lower flux threshold, which we set as \SI{10}{\milli \jansky \per \beam}.

\section{Stacking}

\subsection{Selection of halo pairs}

\begin{figure}
    \centering
    \includegraphics[width=0.9\linewidth, clip, trim={0 0cm 0 1cm}]{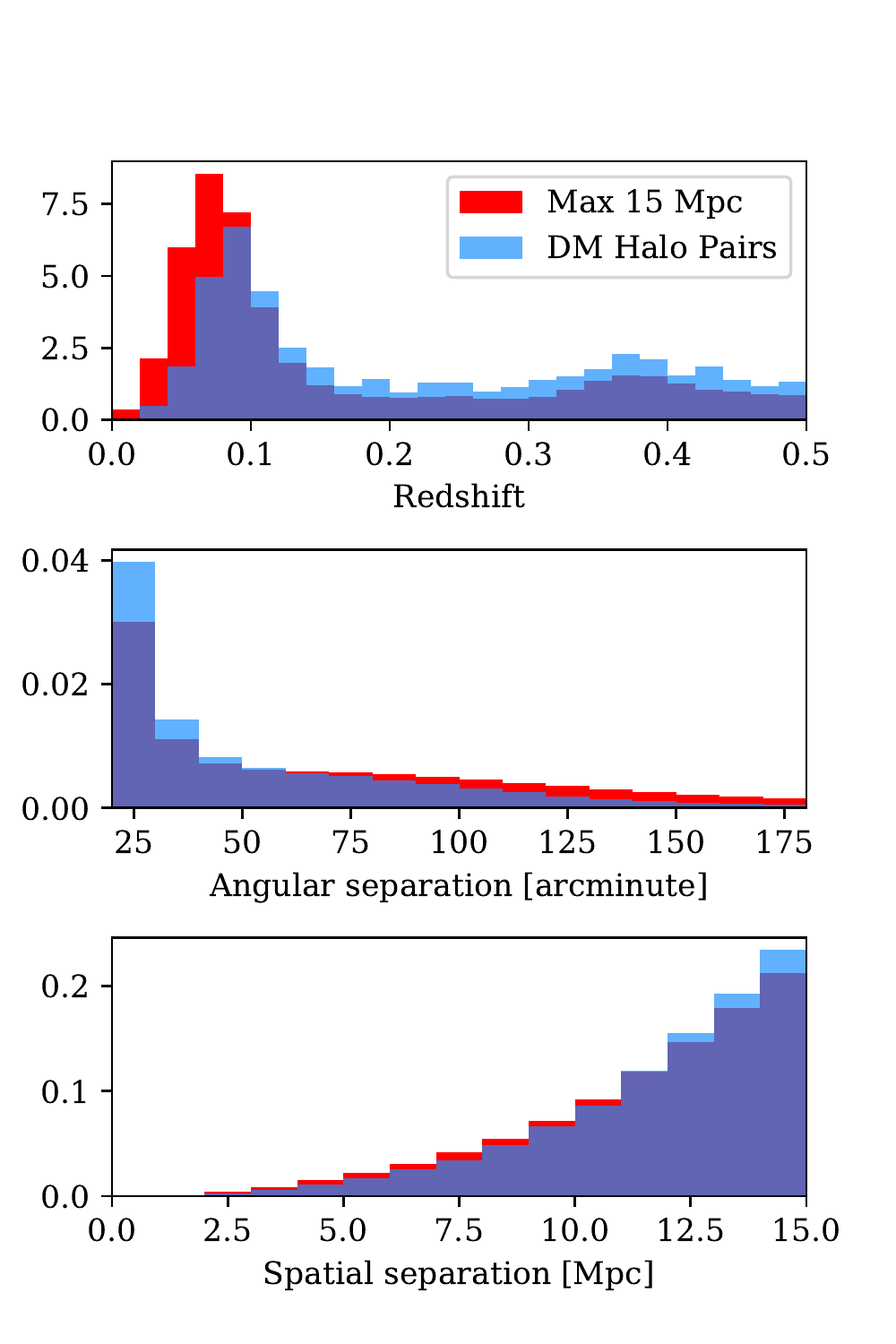}
    \caption{The redshift, angular separation and spatial separation distributions of our catalogue of halo pairs (blue) compared to the `Max 15 Mpc' catalogue of LRG pairs from H2022 (red). Histograms are normalised so as to integrate to unity.}
    \label{fig:halostats}
\end{figure}


Before we can perform stacking, we must first create a catalogue of DM halo pairs. In V2021 and H2022, LRG pairs were drawn from the LRG catalogue constructed by \citet{Lopes2007} if they met the following criteria: where the angular separation of a pair $\theta$ satisfied \SI{20}{\arcminute} $< \theta <$ \SI{180}{\arcminute}; and the physical separation satisfies \SI{1}{\mega \parsec} $< R <$ \SI{15}{\mega \parsec} (comoving).\footnote{The lower threshold on $R$ is not documented in V2021, but was provided in personal communication, as was the use of a comoving distance metric.}

In contrast to V2021 and H2022, where cluster locations were inferred only by using LRGs as proxy, we have perfect knowledge of the DM halos within the simulation. It is therefore possible for us to exhaustively extract all such DM halo pairs that meet this criteria. Doing so, however, would result in a population of DM pairs with notably different distribution of spatial and angular separation, as well as skewed towards significantly deeper redshifts than used in these prior experiments. In fact, the stacking procedure is relatively sensitive to the distribution of pairs used in the stack, and in particular their angular separation (as this determines the required rescaling of the image during stacking) as well as their redshift (as this will on average determine the flux).

Thus we have chosen to select DM halo pairs that approximate the angular, spatial and redshift distribution LRG pairs found in the `Max 15 Mpc' catalogue from H2022. To do this, we have simply binned this reference catalogue in all three dimensions so as to roughly generate a probability distribution. From our exhaustive list of DM halo pairs, mass limited to $M_{\odot} > 10^{12.5}$, we have extracted, without replacement, 70,000 DM halo pairs across the ten realizations based on this probability distribution. In \autoref{fig:halostats}, we plot the  redshift, angular separation and spatial separation of our 70,000 DM halo pairs in (blue) in comparison to those found in the Max 15 Mpc catalogue (red). This process does a reasonably good job at replicating the LRG pair distribution, including the twin peaks found in the redshift distribution that is artifact found in the original LRG catalogue of \citet{Lopes2007}. There is, however, some small deviation between the two distributions, and this has arisen where our own DM halo catalogue was exhausted for certain combinations of parameters.

\subsection{Stacking and modelling methodology}
\label{sec:stackandmodel}

Stacking proceeds nearly identically as described in V2021, and using the same code as used in H2022. We briefly summarize the process here. For a given catalogue, we first identify a maximum scaling factor and associated maximum pixel length. Then for each DM halo pair, we rescale the image, strictly larger, such that the pixel distance between the DM halo pair exactly matches the maximum pixel length. This rescaled image is then rotated so that the line spanning the DM halo pair is rotated to horizontal and such that each DM halo is aligned to normalised coordinates at $(-1, 0)$ and $(+1, 0)$, before the image is then cropped and stacked onto previously processed cutouts. A respective weight map is also created, set to one for valid cutout pixels or to zero for invalid or out of bounds pixels, and this is similarly stacked. Finally, when all halos have been processed and stacked, the stacked DM halo pairs are divided by the stacked weight maps.

Once stacking is completed, as with both V2021 and H2022, we observe clearly discernible peaks of emission at the stacked DM halo centres, at $(-1, 0)$ and $(+1, 0)$. V2021 made the assumption that the majority of this emission is attributable to cluster emission such as AGN and SFG populations, radio halos, and other radio emission processes typically found in cluster environments. Crucially, they also made the assumption that this cluster emission should be, on average, radially symmetric, as opposed to the weaker filamentary emission which should only be present between the pair of clusters. Thus V2021 constructed a radio profile model for this core emission based on the exterior radial profile around each peak, and which they could later subtract so as to reveal excess intracluster emission. To do this, the profile was calculated by finding the radial average of emission around each of the central peaks, but calculated only across a \SI{180}{\degree} sweep strictly `behind' the intracluster region (i.e.\@ for $x < -1$ or $x > 1$, for each peak respectively).

H2022 implemented an identical modelling procedure, and verified this process on synthetic cosmic web images. We have used their modelling code without modification.

\subsection{Noise estimation}

\begin{figure}
    \centering
    \includegraphics[width=\linewidth]{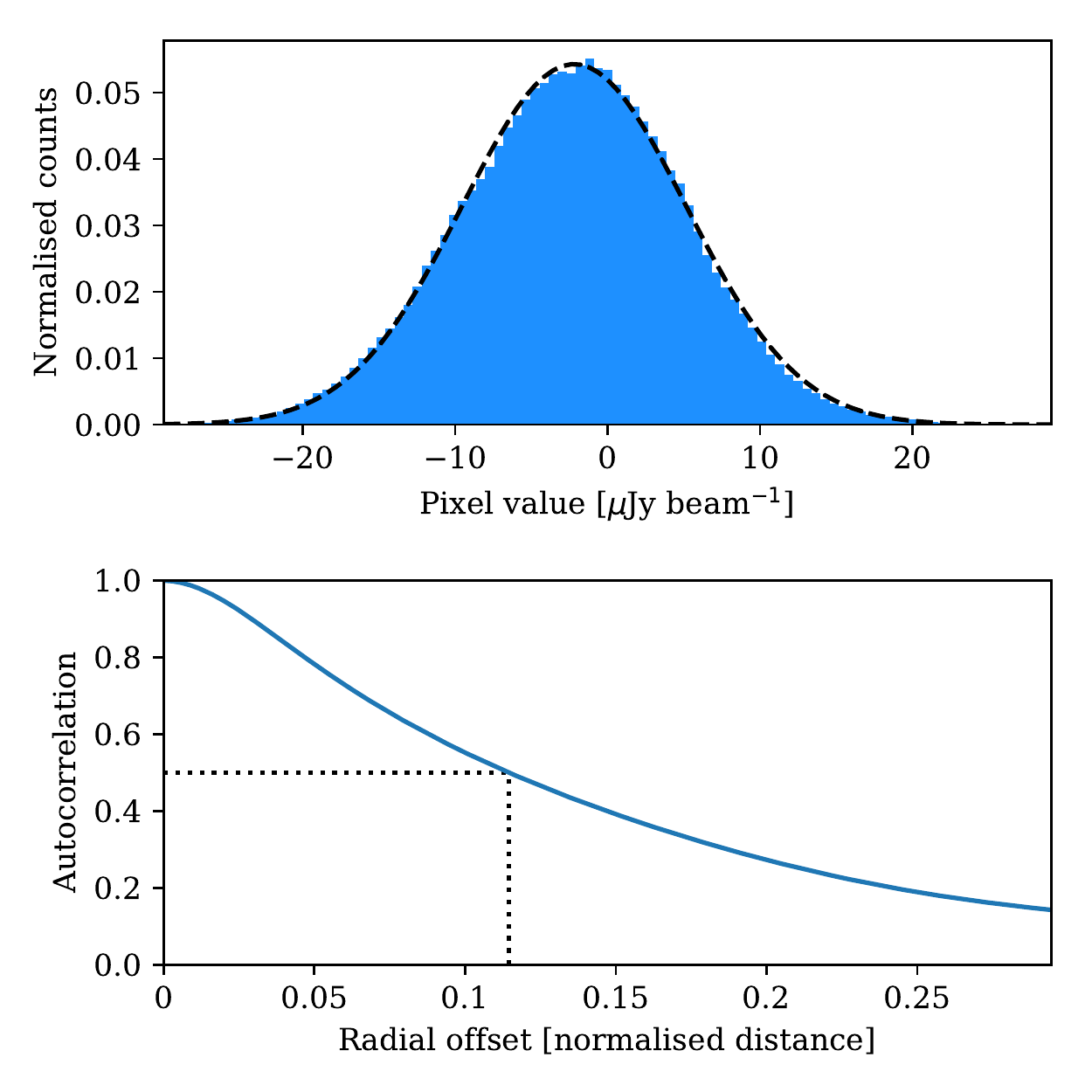}
    \caption{The noise properties of the stacked image for all combined observables with Gaussian beam, calculated across the model-subtracted image, excluding circular regions around the peaks of radius $r = 0.2$ and the inner region boundef by $-1 < x < 1$ and $-1 < y< 1$. \textit{Upper panel:} The distribution of the pixel values (blue) and a fitted Gaussian (dashed black line) showing the stacked noise is approximately normal in distribution. \textit{Lower panel:} The radial autocorrelation of this region, showing the pixel to pixel correlation. The dotted black line indicates the half width, half maximum of this autocorrelation.}
    \label{fig:noisestats}
\end{figure}

In V2021 and H2022, the final stacked images were shown to be approximately normally distributed. This arises quite naturally as the original images are themselves dominated by noise arising from system noise, sky temperature and sidelobe confusion. In this case, we do not start off with noisy images: our simulated fields do not have any injected noise. Nonetheless, we find that after stacking, our stacked images have the appearance of noise that arises from variably sized cutouts of real emission features being scaled and rotated many thousands of times. To measure the typical distribution of pixels within our stacks, we construct an area within our model-subtracted stacks sufficiently far from either peak as well as the intracluster region: this excluded region is therefore the union of the region bounded by $-1 < x < 1$ and $-1 < y < 1$, as well as the regions with radius $r < 0.2$ of either peak at $(-1, 0)$ or $(1, 0)$. In \autoref{fig:noisestats}, we show in the upper panel the distribution of these pixels from the combined (AGN, SFG, cosmic web) stack. In fact, this distribution of pixels is still well approximated by a Gaussian distribution, as illustrated by the close fit of the Gaussian envelope parameterised by $\sigma = 7.34$, and we shall proceed to quantify the noise herein simply by the standard deviation.

Additionally, we note that this noise is spatially correlated across the stacked images. The size of this spatial correlation is a function both of the original resolution of the fields prior to stacking combined with magnitude of rescaling operations during stacking. Following H2022, we measure the resulting `effective resolution' of the stacked image by way of autocorrelating the image. For example, in the lower panel of \autoref{fig:noisestats} we show the autocorrelation of the combined image, using the same area over which we made the earlier noise calculation. From this plot we can read off a half width at half maximum value of 0.11 for the autocorrelation, which corresponds to an effective resolution of the original stacked image having a full width at half maximum (FWHM) of 0.16.\footnote{The effective resolution of the original image is related to the HWHM of the autocorrelation by the relation $\theta_\text{FWHM} = \sqrt{2} \cdot \theta_\text{HWHM}$. This derives from the fact that the autocorrelation of a Gaussian function produces another Gaussian, with its width increased by a factor of $\sqrt{2}$.} Emission structures on scales larger than this effective resolution are likely therefore to be extended.

\section{Results \& Discussion}

\begin{figure*}
    \centering
    \begin{subfigure}{\linewidth}
        \centering
        \includegraphics[width=0.7\linewidth,clip,trim={2cm 1cm 0cm 2cm}]{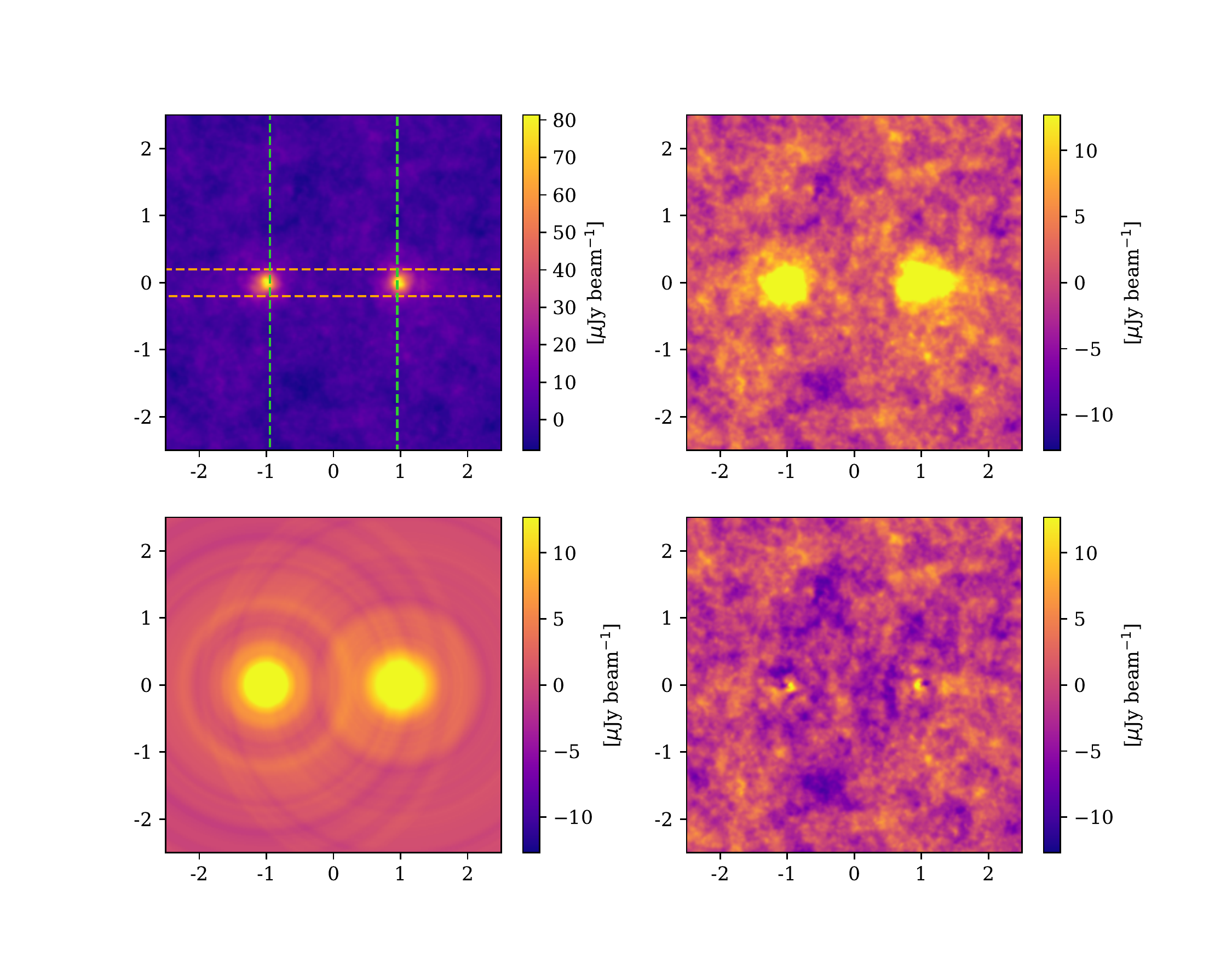}
        \caption{\textit{Top left:} The original mean stack image, with overlays indicating the region over which the transverse mean (dashed orange horizontal lines) and longitudinal mean (dashed green vertical lines) are calculated. \textit{Top right:} The mean stacked image with the colour scale reduced to $\pm 5 \sigma$ to emphasise the noise. \textit{Bottom left:} The model image, on the same colour scale. \textit{Bottom right:} The residual stack after model subtraction, with the colour scale set to $\pm 5 \sigma$.}
        \label{fig:webstacka}
    \end{subfigure}
    \begin{subfigure}{\linewidth}
        \centering
        \includegraphics[width=0.8\linewidth,clip,trim={0 0 0 -0.5cm}]{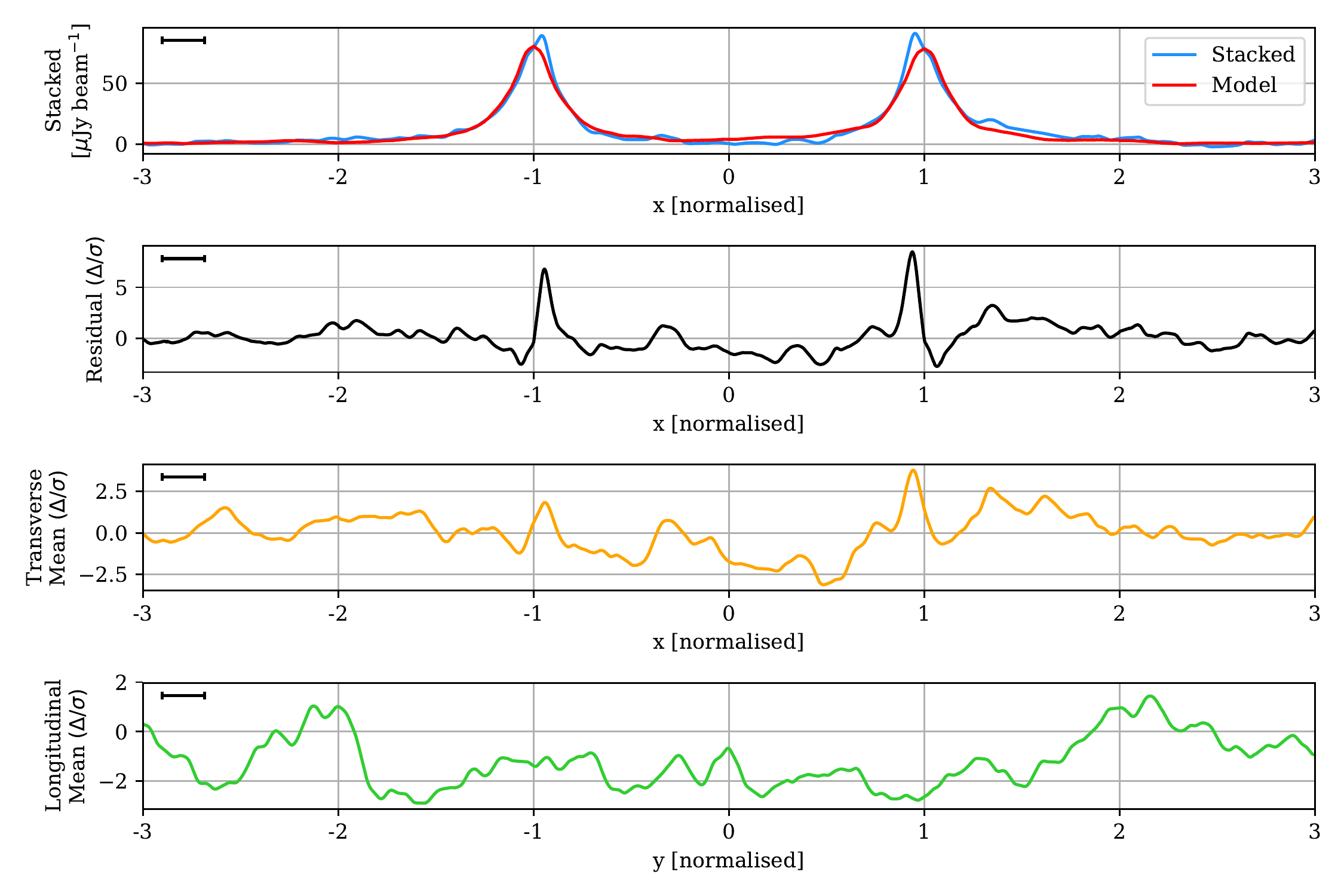}
        \caption{\textit{One:} The one-dimensional profile along $y = 0$ for both the stacked image (blue) and the model (red). \textit{Two:} The one-dimensional profile along $y = 0$ of the residual stack, renormalised to the estimated map noise. \textit{Three:} The transverse mean along the region $-0.2 < y < 0.2$ of the residual stack, renormalised to the estimated map noise. \textit{Four:} The longitudinal mean along the region $-0.95 < x < 0.95$ of the residual stack, renormalised to the estimated map noise. The black rule in the top left shows the FHWM of the effective resolution.}
        \label{fig:webstackb}
    \end{subfigure}

    \caption{The synchrotron cosmic web stack, with estimated noise \SI{2.5}{\micro \jansky \per \beam} and effective resolution 0.21. The left peak has a maximum of \SI{89.46}{\micro \jansky \per \beam} and a FWHM of 0.27; the right peaks at \SI{91.1}{\micro \jansky \per \beam} and has a FWHM of 0.28.}
    \label{fig:webstack}
\end{figure*}

\begin{figure*}
    \centering
    \begin{subfigure}{\linewidth}
        \centering
        \includegraphics[width=0.7\linewidth,clip,trim={2cm 1cm 0cm 2cm}]{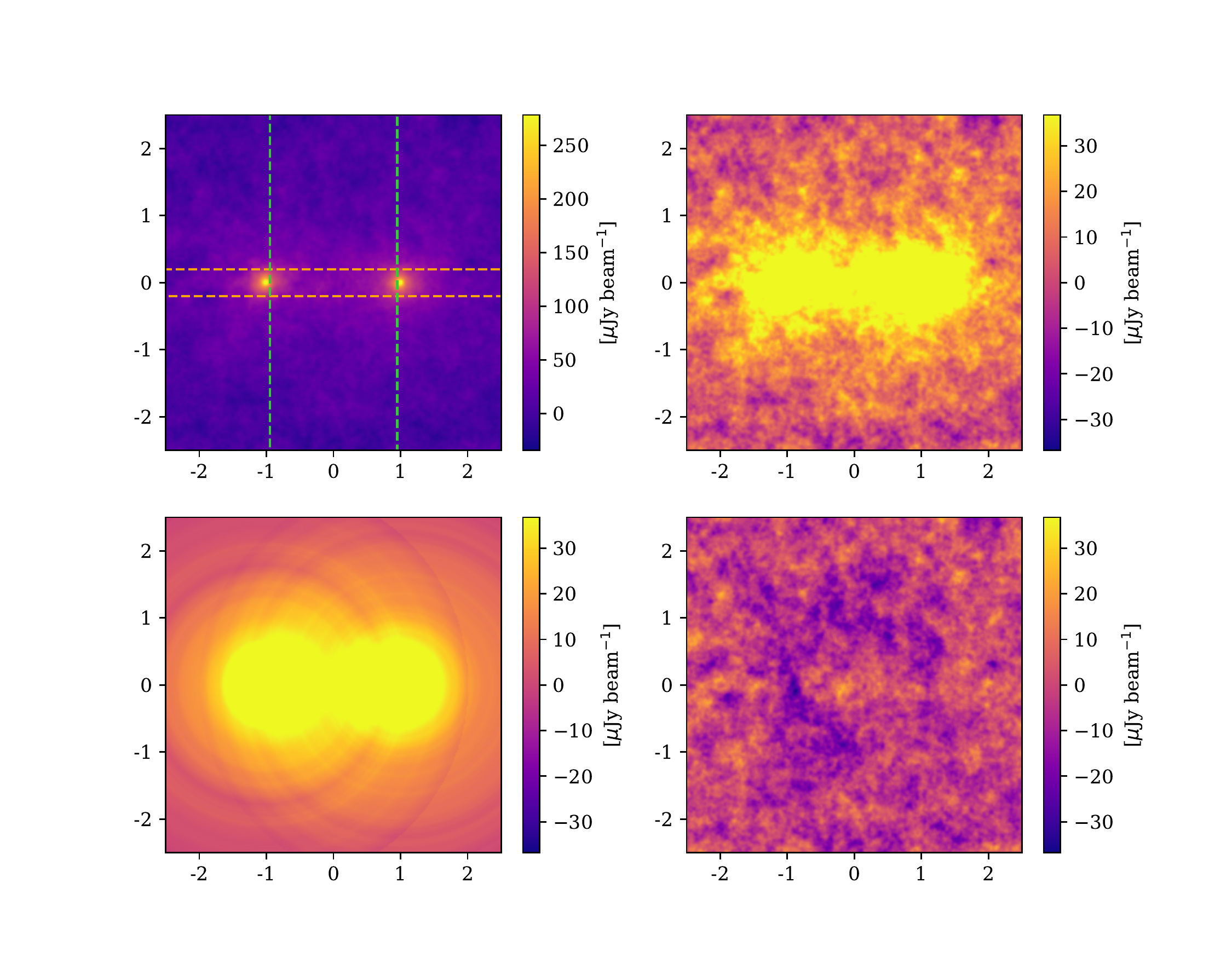}
        \caption{\textit{Top left:} The original mean stack image, with overlays indicating the region over which the transverse mean (dashed orange horizontal lines) and longitudinal mean (dashed green vertical lines) are calculated. \textit{Top right:} The mean stacked image with the colour scale reduced to $\pm 5 \sigma$ to emphasise the noise. \textit{Bottom left:} The model image, on the same colour scale. \textit{Bottom right:} The residual stack after model subtraction, with the colour scale set to $\pm 5 \sigma$.}
        \label{fig:combinedstacka}
    \end{subfigure}
    \begin{subfigure}{\linewidth}
        \centering
        \includegraphics[width=0.8\linewidth,clip,trim={0 0 0 -0.5cm}]{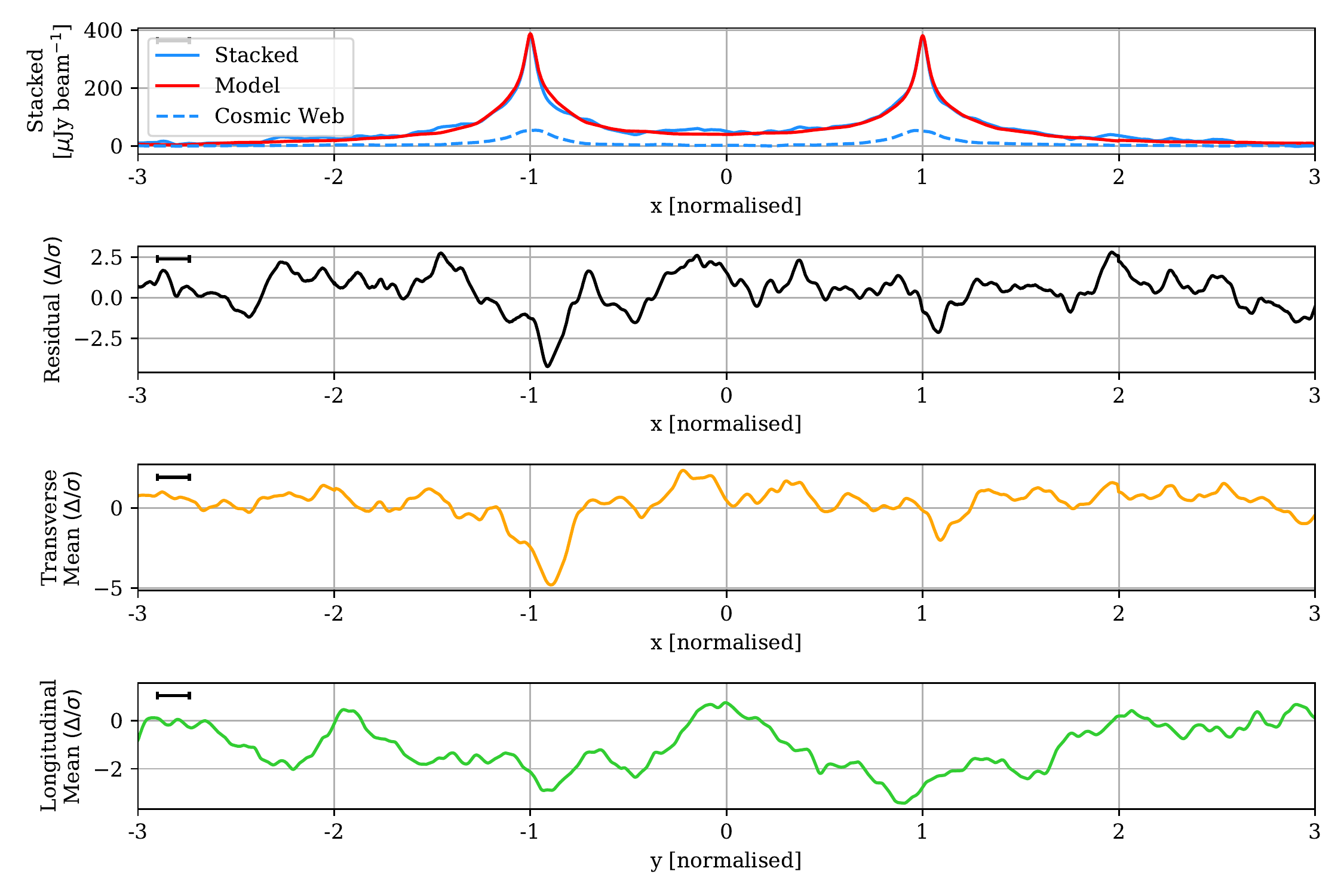}
        \caption{\textit{One:} The one-dimensional profile along $y = 0$ for the stacked image (blue), the model (red) and the synchrotron cosmic web component (dashed blue). \textit{Two:} The one-dimensional profile along $y = 0$ of the residual stack, renormalised to the estimated map noise. \textit{Three:} The transverse mean along the region $-0.2 < y < 0.2$ of the residual stack, renormalised to the estimated map noise. \textit{Four:} The longitudinal mean along the region $-0.95 < x < 0.95$ of the residual stack, renormalised to the estimated map noise. The black rule in the top left shows the FHWM of the effective resolution.}
        \label{fig:combinedstackb}
    \end{subfigure}

    \caption{The combined (AGN, SFG, cosmic web) stack, with estimated noise \SI{7.3}{\micro \jansky \per \beam} and effective resolution 0.16. The left peak has a maximum of \SI{374.9}{\micro \jansky \per \beam} and a FWHM of 0.16; the right peaks at \SI{371.7}{\micro \jansky \per \beam} and has a FWHM of 0.14.}
    \label{fig:combinedstack}
\end{figure*}

We begin by presenting the stacked results for the synchrotron cosmic web alone, which we isolate here to focus on its unique signature. These results stack 70,000 DM halo pairs across all ten realisations, and in this case we have not undertaken any point source subtraction. In \autoref{fig:webstacka}, the upper plots show the stacked image on two different colour scales, in which the upper right is scaled to emphasise the noise at \SI{2.5}{\micro \jansky \per \beam}. We observe peaks centred approximately at $(-1, 0)$ and $(1, 0)$, and having maximum values of approximately \SI{90}{\micro \jansky \per \beam} and FWHM widths of about 0.28. In the lower left plot, we show the model constructed as per \autoref{sec:stackandmodel}, and in the lower right panel we show the stack after model subtraction. In \autoref{fig:webstackb}, we show one-dimensional profiles showing: at the top, the profile along $y = 0$ of the original stacked image compared to the model; in the second row, the profile along $y = 0$ of the residual image; in the third row, the tranverse mean in the region $-0.2 < y < 0.2$ as a function of $x$; and finally at the bottom, the longitudinal mean in the region $-0.95 < x < 0.95$ as a function of $y$. We apply the transverse mean in an attempt to draw out any faint but wide signals along the intracluster region; whilst the longitundinal mean attempts to bring out any faint signals the might span the length of the intracluster region.

The peaks that we observe are primarily the result of radio relic-like shocks surrounding the clusters and galaxy groups of the FIGARO simulation. As noted in \citet{Hodgson2021}, some 96\% of the radio power of cosmic web emission in FIGARO was located about the 100 most massive DM halos, within the spherical regions $r < 1.5 r_{200}$ (where $r_{200}$ is the $M_{\odot} = 200 \rho_c$ virial radius of the cluster). This radio power is generated by stationary accretion shocks that are morpologically akin to radio relics, often appearing as thin, elongated arcs that trace the shock front.

Outside of cluster peripheries, as noted in \citet{Hodgson2021}, the emission that traces the filaments proper is orders of magnitude fainter. Indeed, examining \autoref{fig:webstack} closely, we make no detection of faint filament emission along the intracluster region in our stack, nor in the one dimensional profile, or the transverse and longitudinal means. Our modelling suggests that in a stacking experiment like this, the intracluster region will be devoid of any detectable emission attributable to the cosmic web.

We do, however, note a curious detail about the peaks: they are asymmetric about their respective centres. Both the left and right peaks are actually shifted slightly inward, and this is made especially clear when we construct the model, which assumes radial symmetry, as seen in the top panel of \autoref{fig:webstackb}. As a result, the residual image in the lower right of \autoref{fig:webstackb} shows peaks of emission left behind, and this is illustrated further in the one dimensional profile of the residual image in panel two of \autoref{fig:webstackb}, where we see peaks slightly interior to the halo pair reaching more than $5 \sigma$. We expand upon this feature shortly.

Next, we turn to the results in \autoref{fig:combinedstack} of stacking the combined fields, which incorporate all of the AGN, SFG and cosmic web emission in one. This combined field has had bright sources subtracted, primarily AGN and SFG sources, by cleaning down to \SI{10}{\milli \jansky \per \beam}. These additional sources increase the stacked noise which is about three times higher at \SI{7.3}{\micro \jansky \per \beam}. As previously, we note in these stacks bright peaks of emission at x = $\{-1, +1\}$, however in this case they are much brighter, at approximately \SI{373}{\micro \jansky \per \beam}, as well as much narrower, having a FWHM width of about 0.15. The effective resolution of the map is 0.16, meaning these peaks are essentially point-like, and are dominated by the compact emission of the AGN and SFG population. Note that the peak width is much narrower than observed in V2021, however similar to that found in H2022 and consistent with what would be expected after stacking a principally unresolved population of sources. Note that the maximum peak values are primarily simply a function of the cleaning threshold.

Any cosmic web emission present in the stacked map is dominated by these AGN and SFG populations. In the top panel of \autoref{fig:combinedstackb}, we show the one-dimensional profile of the combined stack in blue, and in dashed blue the cosmic web contribution (after cleaning) towards this peak. Whilst the cosmic web contribution is extended, having a FWHM of 0.28, this is lost in the combined peaks, which as we noted appear point like. Moreover, the slight asymmetry of the cosmic web contribution is lost amongst the added noise of the AGN and SFG populations. The one dimensional profile of the residual stack after model subtraction now shows no evidence of the peaks slightly interior to the halo pair that we observed earlier, and in fact there is a negative peak at $x = -0.92$; since this depression is not mirrored on both sides of the stack about the centre at $x = 0$, we attribute this to stack noise.

The stated temperature of the filaments detected in V2021 was \SI{100(40)}{\milli \kelvin} at \SI{150}{\mega \hertz}. This value is equivalent to a flux density in our stacks of \SI{51}{\micro \jansky \per \beam}, which would amount to an approximately $7 \sigma$ signal, even in the noisier combined stack. It is clear that these simulated stacks do not support either the magnitude or location of the excess emission as detected in V2021.

\subsection{The ubiquity of `relic'-like shocks}

\begin{figure}
    \centering
    \includegraphics[width=0.8\linewidth]{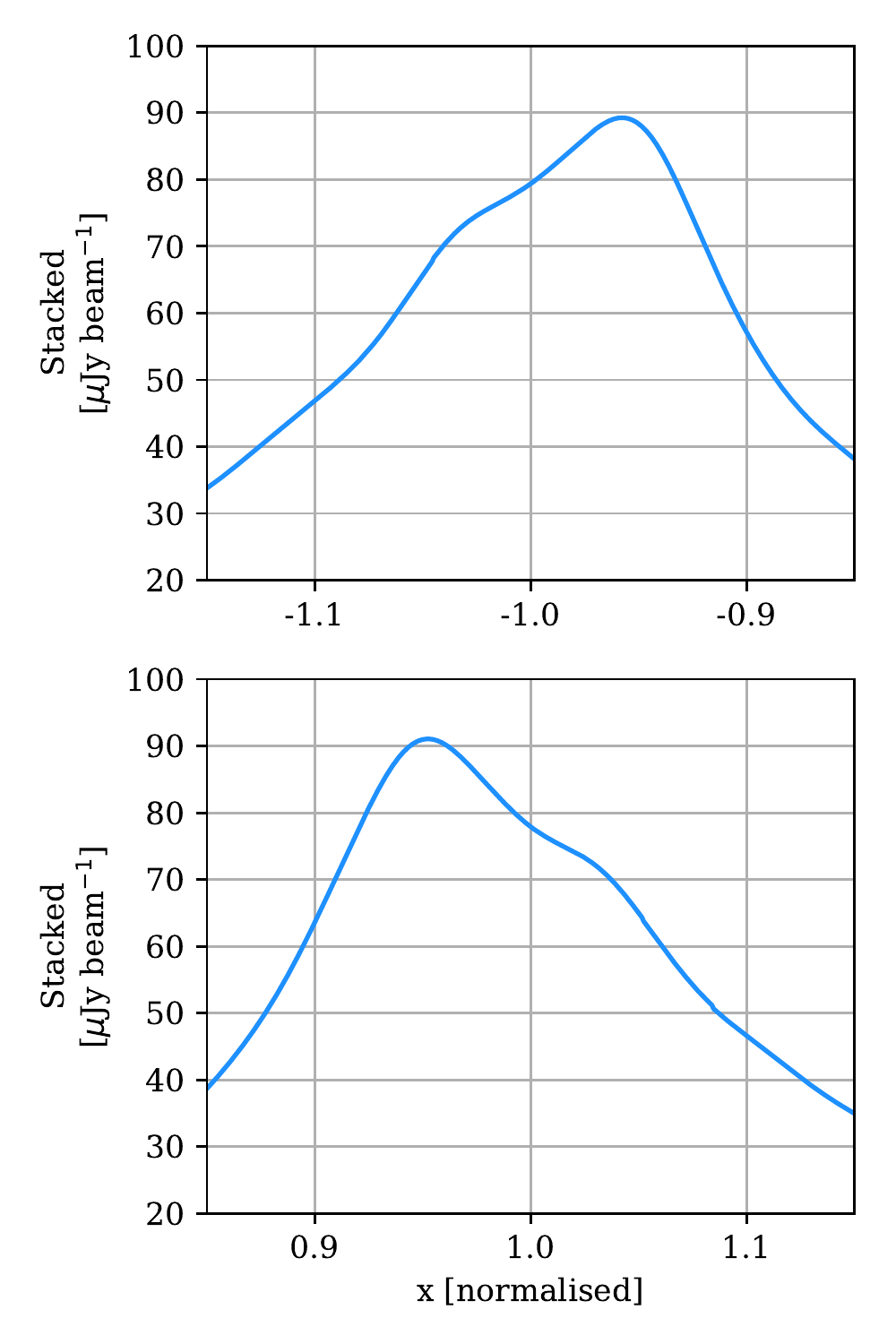}
    \caption{Zoomed plots of the peaks at $x = \{-1, 1\}$ in the  one-dimensional profile of the stacked synchrotron cosmic web along $y = 0$ from Figure \ref{fig:combinedstackb}. Note the profile peaks are interior to the intracluster region, at approximately $x = \pm 0.95$, as well as the inflection point at around $x = \pm 1$.}
    \label{fig:peakcloseup}
\end{figure}

\begin{figure}
    \centering
    \includegraphics[width=\linewidth]{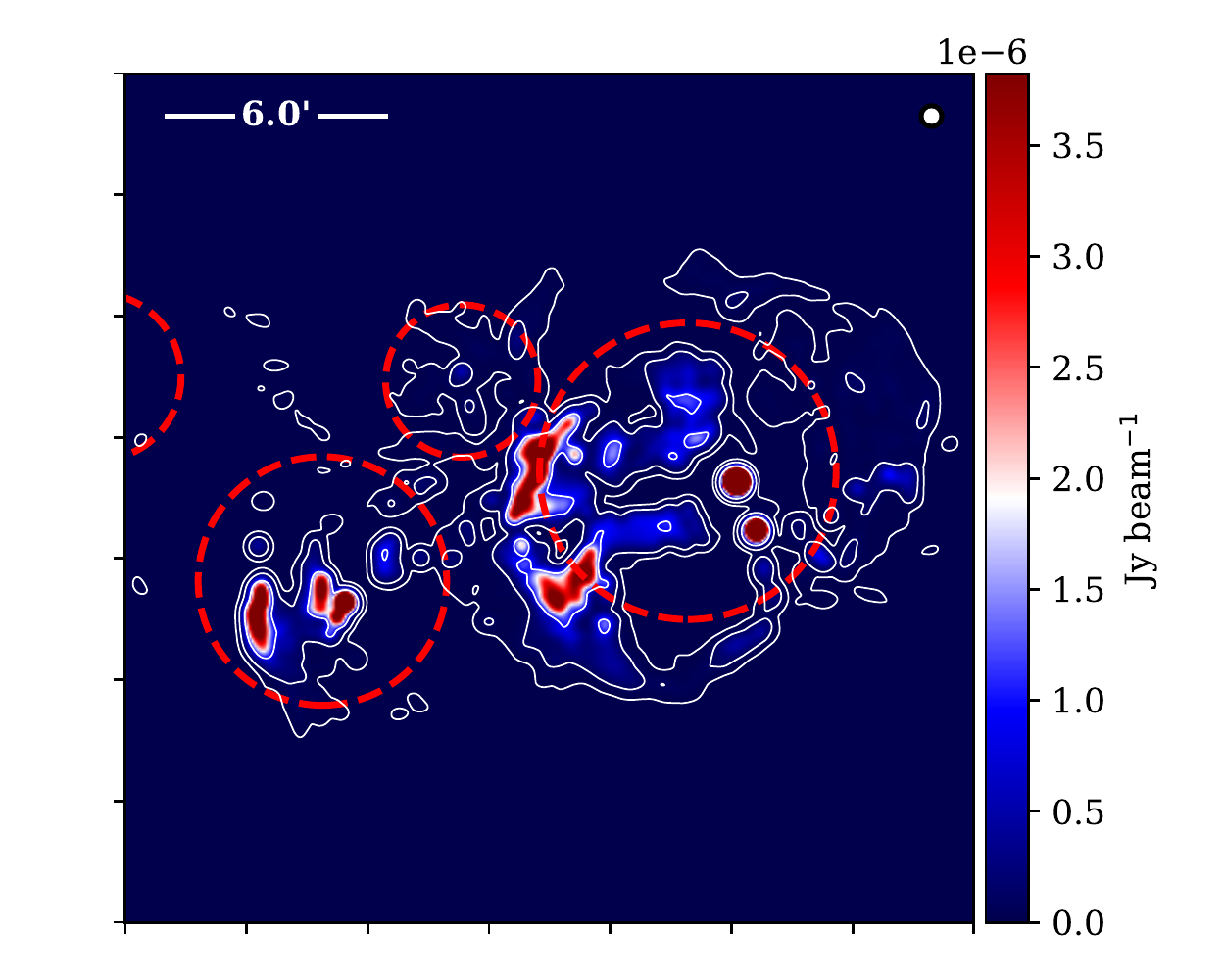}
    \caption{An example from FIGARO of asymmetric accretion-driven shocks about the periphery of interacting clusters. We observe three close-proximity clusters in the redshift range $0.15 < z < 0.2$, with approximate virial radii ($r_{200}$) indicated by dashed red lines. The contours show the integrated radio emission along the line of sight observed at \SI{150}{\mega \hertz} and at a FWHM beam resolution of \SI{20}{\arcminute}. The rightmost cluster displays a pair of parenthetical arcs of emission about its centre, however with the interior arc significantly brighter. Whilst the projection makes some emission appear centrally located within cluster interiors, all emission is located at a radius from cluster centres of at least $r > 0.8 \cdot r_{200}$.}
    \label{fig:interesting}
\end{figure}

One of the key observations made about FIGARO in \citet{Hodgson2021} was the degree to which the distribution of emission sources within the synchrotron cosmic web differed from prior expectations. It had been expected that emission structures traced the underlying mass of the filaments. Indeed, the \citet{Vazza2019} MHD simulation showed that the X-ray emission does trace this mass distribution. However, for the synchrotron component of the cosmic web, \citet{Hodgson2021} noted a large population of emission structures in the spherical shell of DM halos, in the range $0.75 \cdot r_{200} < r < 1.5 \cdot r_{200}$ where $r_{200}$ is the virial radius of the DM halo. These emission structures were morphologically similar to radio relics, most often tracing parenthetical arcs on opposing sides of the DM halo core. Outside of these regions, the filaments hosted emission orders of magnitude fainter. However, the mechanism for these shocks differs from radio relics: whilst relics are predominantly driven by cluster-scale merger events, the majority of these shocks are driven by strong accretion shocks at the virial boundary of overdense regions, forming stationary emission structures.

As we have noted, the cosmic web stacks display a unique asymmetry about the stacking centres, and instead peak slightly interior to halo pairs, at around $x = \pm 0.95$. In \autoref{fig:peakcloseup} we show zoomed plots of the peaks of the stacked cosmic web, showing the one dimensional profile along $y = 0$ around each peak. In addition to the peaks positioned interior to the intracluster region, we also note in these zoomed plots the presence of a second smaller peak at around $x = \pm 1.05$, and point of inflection in the vicinity of $x = \pm 1$.

These features can be attributed to asymmetric accretion shocks about cluster peripheries, combined with projection effects. If we consider a double `relic' cluster---two arcs of emission about an otherwise radio-quiet cluster core---then depending on the angle of observation this system will either appear as two separate peaks of emission, a single peak of emission where both the relic pair and the DM halo core are aligned along the line of sight, or some intermediary combination. When integrated over many such systems in the stacked image, these projection effects will contribute both to an emission peak at $x = \{-1, 1\}$ as well as additional emission immediately surrounding. Moreover, the asymmetry of the double-peak structure indicates that on average that the interior relic is more emissive (see \autoref{fig:interesting} for one such example). We can infer that the interior shock is subject to some combination of stronger shocks, a denser or hotter electron environment, or stronger magnetic fields; these are reasonable effects where the interior environment is slightly more compressed than the exterior.

\subsection{Filamentary emission}

\begin{figure}
    \centering
    \includegraphics[width=\linewidth,clip,trim={0cm 0cm 1cm 0cm}]{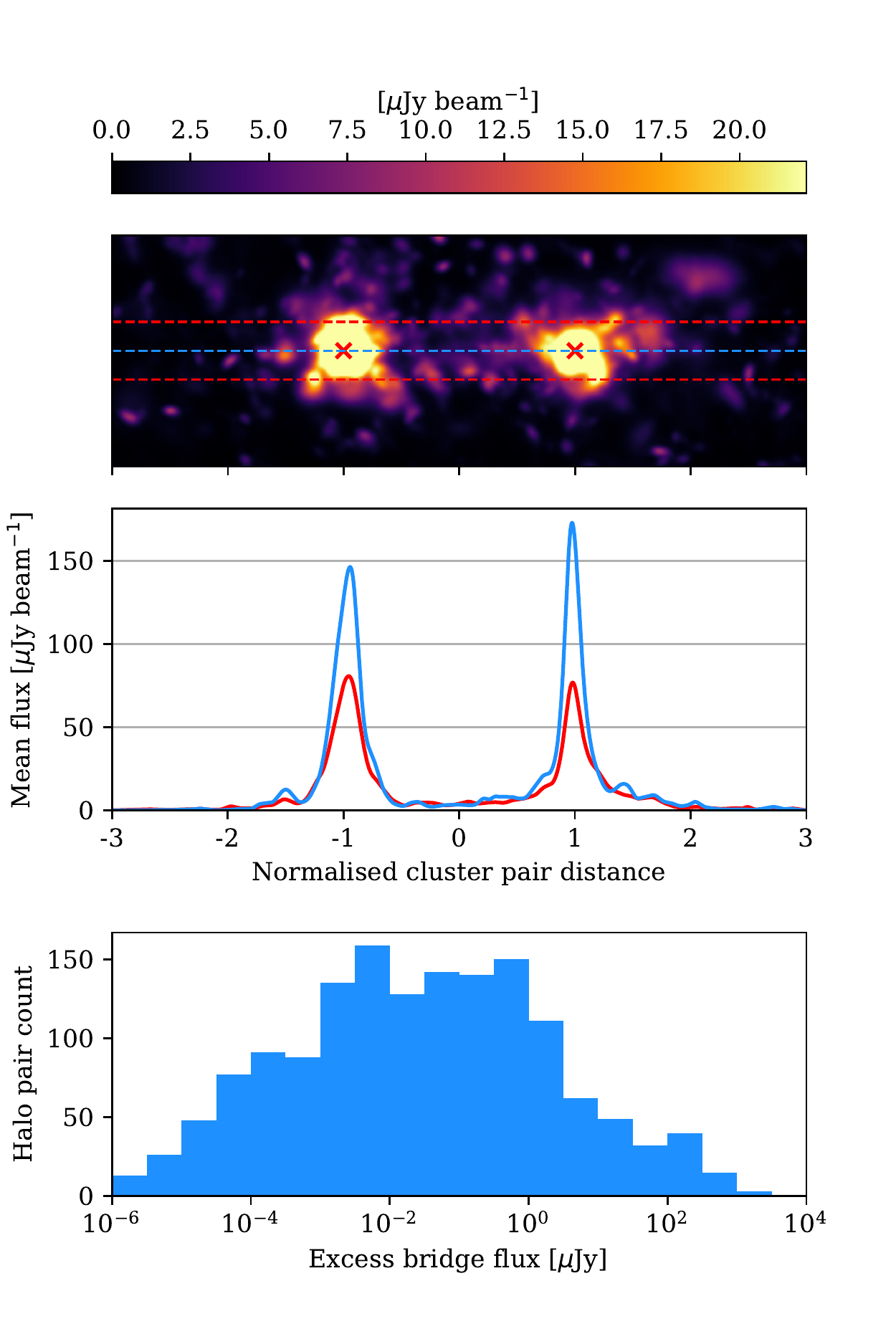}
        \caption{Stacked cosmic web emission between DM halo pairs (with $M >$~\SI{E12.5}{\solarmass}, \SI{1}{\mega \parsec}~$< r <$~\SI{15}{\mega \parsec}) within the original simulation volume (snapshots 166 \& 188) from \citet{Vazza2019}, set at a redshift $z = 0.14$ and an observing frequency of \SI{118}{\mega \hertz}. Crucially, this stack isolates the halo pair from all foreground and background emission. \textit{Upper panel:} The stacked image of halo pairs, scaled such that the colour saturates at the 99th percentile pixel. \textit{Centre panel:} The values measured between the two halo peaks (blue) along the line indicated in dashed blue; and the mean values (red) calculated in the region between the dashed red lines. \textit{Lower panel:} The distribution of excess intracluster emission for individual halo pairs, showing that the majority of the bridge excess is the result of just a small handful of pairs.}
    \label{fig:3Dstack}
\end{figure}

We have not been able to detect excess filamentary emission in our stacks that is detectable above the stacking noise. However, we could suppress this stacking noise if we could exclude all foreground and background emission sources, and stack the DM halo pairs in isolation. V2021 attempted a similar calculation on the same simulation data from \citet{Vazza2019} and we reproduce this exercise here, however differing in a few key respects: we extract, isolate and rotate the halo pair from within the three-dimensional volume, whereas they used only a flattened map; we use the simulation emission calibrated as per \citet{Hodgson2021}, resulting in an approximately six-fold increase in total emission across the volume; and in our analysis we also calculate the `background' emission against which any excess would be measured. The latter difference is important, as the model construction method used here and in V2021 measures the \textit{excess} emission with respect to the background, not the true strength of the filamentary emission itself.

Our method proceeded by extracting all halos with $M >$~\SI{E12.5}{\solarmass} from the two nearest snapshots of the original MHD simulation, of which there are 931, and finding all halo pairs with comoving distance in the range \SI{1}{\mega \parsec}~$< r <$~\SI{15}{\mega \parsec} and angular separation \SI{20}{\arcminute}~$< \theta <$~\SI{180}{\arcminute} when placed at a redshift of $z = 0.14$, of which there are 4919 pairs. For each pair, we have rotated and scaled the volume to align the pair at normalised coordinates $x = \{-1, 1\}$ and removed all emission in the foreground or background where $\abs{z} > 0.5$. To match the method used in V2021, all emission was placed at a redshift of $z = 0.14$, however our method differs in that we use the PSF used throughout this paper and that we set the observing frequency to \SI{150}{\mega \hertz}.

In \autoref{fig:3Dstack} we present the results of this isolated stack, showing the mean flux density of the full stack in the upper panel, and in the centre we show both the one-dimensional strip of values of this stack (blue) as well as the mean value across the region $-0.25 < y < 0.25$ (red). The peaks near $x = \{-1, 1\}$ have a mean value of \SI{162}{\micro \jansky \per \beam}, with both peaks offset towards the interior. \footnote{The cosmic web peaks here are brighter than those stated previously in our stacked cosmic web. This discrepancy arises as the model construction process used in stacking subtracts away the mean background emission (produced from unrelated foreground and background emission sources) and so the peak values given there are actually only the excess peak emission above this background. In the case of the cosmic web stack, this background emission was \SI{65.5}{\micro \jansky \per \beam}.} The mean value in the intracluster region spanning $-0.5 < x < 0.5$ and $-0.25 < y < 0.25$ is \SI{4.4}{\micro \jansky \per \beam} or equivalently \SI{8.7}{\milli \kelvin}. This value is in comparison to the mean background emission of \SI{3.0}{\micro \jansky \per \beam}, which we have calculated as the mean value across the full exterior \SI{180}{\degree} sweep about each peak, in the radial range $0.5 < r < 1.5$. The excess filamentary emission with respect to this background is therefore \SI{1.4}{\micro \jansky \per \beam} or equivalently \SI{2.7}{\milli \kelvin}.

In the lower panel of \autoref{fig:3Dstack}, we also present the distribution of individual halo pair contributions to the mean intracluster excess. Clearly, a small fraction of halo pairs are responsible for the bulk of the measured excess emission along the bridge, with the majority providing negligible signal. Note also that the turnover at lower excess fluxes is an artificial result of the minimum mass threshold; modification of this lower threshold moves the location of the peak in this distribution.

This is an artificial exercise, however it does tell us a couple of things. Firstly, that there is indeed a small intracluster excess present in the underlying simulation; secondly, that this emission is some two orders of magnitude lower than the cosmic web emission around the periphery of clusters and groups; and thirdly, that a small handful of intracuster pairs are responsible for the bulk of the emission. Detecting this excess with stacking, however, would require vastly more pairs of DM halos and observed sky area to sufficiently suppress the stacking noise. For example, we would need to increase the number of stacked DM halo pairs in our simulated stacks from 70,000 to approximately 17 million to allow a $3\sigma$ detection of this excess filamentary emission.

\subsection{Limitations on the current study}

The conflicting findings between V2021 and H2022 remain unresolved. Whilst this simulated stacking exercise does not support the findings of V2021, its important to note some key limitations. In this section, we discuss some of the salient limitations of the underlying MHD simulation as well as explore possibilities where the simulated synchrotron emission along filaments might be amplified. 

The first important caveat of note is that in the densest, most massive parts of the MHD simulation volume, there is good reason to believe synchrotron emission is underestimated. This underestimation arises from the simulation ignoring the role of fossil electrons---electrons that have been previously accelerated either by AGN or historic large-scale accretion shocks---in increasing the acceleration efficiency of shocks. AGN were not modelled as part of the original MHD simulation upon which FIGARO is based, nor were the accumulated effects of earlier epoch shocks; instead, the electron population was always assumed to be at thermal equilibrium. The result of this is to underestimate synchrotron emission in dense environments, especially cluster interiors where fossil electrons can reasonably be expected to survive hundreds of millions of years \citep[e.g.][]{Hodgson2021Jelly}. In the development of FIGARO, we attempted to mitigate this by calibrating the simulated relics to match the known radio relic population \citep{Nuza2012}: dense regions subject to weak shocks had their electron acceleration efficiency artificially increased to $10^{-2}$. Beyond these dense cluster environments, however, the origin, prevalence and lifetimes of fossil electrons in cluster peripheries and in intracluster environments is poorly understood. For example, in their discovery of a ridge of radio emission between merging clusters Abell 399 and 401, \citet{Govoni2019} noted that the observed radio emission was three orders of magnitude brighter than from similar ENZO-based simulations. If, however, the ridge was filled with a population of electrons at energies $\gamma \gtrapprox 1000$, it was possible to boost their simulation emission to match the observation. They did not attempt to explain the origin of this hypothetical fossil electron population, or whether it was especially plausible.\footnote{\citet{Brunetti2020} suggest instead that turbulent Fermi II processes could account for the emission.} Whilst Abell 399 and 401 were separated by \SI{3}{\mega \parsec}, in general a high density of energetic fossil electrons ($\gamma \gtrapprox 1000$) is implausible across the kind of large ($\expval{r} = 10$~\SI{}{\mega \parsec}), low density intracluster environments that are typical of the LRG pairs used in V2021. At best, therefore, a large and energetic fossil electron population could be used to boost nearby pairs of clusters, but cannot be used to boost the filament strength in general.

We must also consider the limited volume of our original MHD simulation, at just $100^3$~Mpc$^3$. This volume reproduces only a handful of massive $\sim$\SI{E14}{\solarmass} clusters, whilst the full region over which V2021 and H2022 performed their stack includes numerous clusters on the order of \SI{E15}{\solarmass}. Whilst few in number, these most massive clusters are likely to be outliers in terms of their contribution to cosmic web emission. It is not clear, however, how such clusters would affect the stacked signal, and whether they would contribute substantially to the intracluster region, or instead simply increase the signal at $x = \{-1, 1\}$. The answer to this question must await larger volume MHD simulations.

Another key input to the MHD simulation is the value that was set for the primordial magnetic field. This was set as \SI{0.1}{\nano \gauss}, about a factor of 10 lower than the upper limits derived from cosmic microwave background observations \citep{Planck2016}. In low density environments, dynamo amplification processes are believed to be negligible and observational data has placed tight limits on such effects \citep{OSullivan2020}. The resulting magnetic field strengths in these sparse environments are instead primarily the result of adiabatic gas compression, and the field strengths along filaments are therefore closely related to the primordial field strength. For values of $B \ll B_\text{CMB}$, where $B _\text{CMB}$ is the equivalent magnetic field strength of the CMB and is approximately $3.25 (1 + z)^2$~\SI{}{\micro \gauss}, the synchrotron emission, $S$, along filaments scales as $S \propto B^2$ \citep{Hoeft2007}. Thus, for small increases in the magnetic seeding scenario, we can significantly amplify the cosmic web emission. This `lever', however, operates globally on both the filaments proper as well as the relics, and quickly becomes unphysically bright. As noted, in the construction of FIGARO, we have already calibrated the cosmic web emission using the radio relic population; further amplification would render a large population of previously hidden radio relics now plainly visible, in contradiction to their observed population statistics. In a sense, then, the brightest outliers of the cosmic web provide a relatively tight constraint on any amplification of the primordial magnetic field.

Finally, we can also consider the effect of a non-uniform primordial magnetic field. Whilst the MHD simulation of \citet{Vazza2019} initialised a uniform magnetic field at redshift $z = 45$, it is possible that these primordial fields already had more complex spatial configurations. Such a change could result in the radio signal of the cosmic web becoming increasingly different outside of DM halo interiors. Dedicated simulations, however, are the only way to accurately predict their effects \citep[e.g.][]{Vazza2021}.

\section{Conclusion}

We have reproduced the stacking and modelling methodology used V2021, and more recently in H2022, using the recently published FIGARO simulation. Using FIGARO, we have constructed ten \SI{15 x 15}{\degree} fields of the radio sky at \SI{150}{\mega \hertz} and for redshifts $z < 0.6$, as observed by the MWA Phase I instrument. We have identified 70,000 DM halo pairs, forming a similar sample to that used in H2022, and have stacked these pairs for each of the combined radio sky (AGNs, SFGs, and synchrotron cosmic web) as well as the synchrotron cosmic web in isolation.

Ultimately, we have been unable to reproduce the specific location of excess intracluster emission observed by V2021. Instead, we observe an excess of emission in the stacks as a result of the cosmic web on the immediate interiors of the stacked peaks at $x = \{-1, 1\}$. These peaks are the result of stationary accretion shocks about the periphery of clusters and galaxy groups, and the asymmetry we observe is likely the result of more emissive interior shocks due to compression effects. These cosmic web peaks, however, are dominated by the AGN and SFG populations when we construct the combined stack, and the small asymmetric peak is obscured by the additional stacking noise.

The true intracluster excess emission in the region $-0.5 < x < 0.5$ is consistent with the noise in our stacked images. In a follow up stack, where we have isolated pairs of DM halos from all foreground and background emission to reduce the stacking noise further, we do find that the intracluster bridge has a small excess of emission compared to the background, but that this excess is two orders of magnitude fainter than the cosmic web peaks.

Our stacking experiment here represents a best case scenario: we know the DM halo population perfectly, and there is no noise or errors associated with our observations. Real world observations and stacking experiments instead stack noisy, imperfectly calibrated images, and currently only have LRGs as an heuristic proxy for the location of clusters. Unless some of the noted limitations to the underlying MHD simulation prove to significantly change the results shown here, it would seem stacking LRG pairs is an especially difficult technique for the detection of the synchrotron cosmic web and future detection attempts will need to stack significantly larger areas of the sky.

\begin{acknowledgements}
The authors thank Tessa Vernstrom for providing additional details regarding the methodology used in V2021. F.V. acknowledges financial support from the ERC  Starting Grant ``MAGCOW'', no. 714196. The cosmological simulations on which this work is based have been produced using the ENZO code (\hyperlink{http://enzo-project.org}{http://enzo-project.org}), running on Piz Daint supercomputer at CSCS-ETHZ (Lugano, Switzerland) under project s805 (with F.V. as PI, and the collaboration of C. Gheller and  M. Br\"{u}ggen).  We also acknowledge the usage of online storage tools kindly provided by the INAF Astronomical Archive (IA2) initiative (\hyperlink{http://www.ia2.inaf.it}{http://www.ia2.inaf.it}).
\end{acknowledgements}


\bibliographystyle{pasa-mnras}
\bibliography{main.bib}

\end{document}